\documentclass[aps,prl,floatfix,preprint,longbibliography,groupedaddress,showkeys]{revtex4-1}

\usepackage{tabularx}
\usepackage{graphicx}
\usepackage{natbib}
\usepackage{pifont}
\usepackage{soul}
\usepackage{amsmath}
\usepackage{color}
\usepackage{mathtools}
\usepackage{relsize}
\usepackage{wasysym}
\usepackage[hidelinks=true]{hyperref}

\newcommand{\vol}{\mathop{\ooalign{\hfil$V$\hfil\cr\kern0.08em--\hfil\cr}}\nolimits}

\newcommand{\bnabla}{\boldsymbol{\nabla}}

\newcommand{\obf}[1]{\overline{\boldsymbol{#1}}}
\newcommand{\norm}[1]{\left\lVert#1\right\rVert}
\newcommand{\tm}[1]{\left\langle#1\right\rangle}
\newcommand{\mrm}[1]{\mathrm{#1}}

\begin{document}


\title{Reduced-order modeling of fully turbulent buoyancy-driven flows using the Green's function method}


\author{M. A. Khodkar$^1$}\thanks{mkhodkar@rice.edu}
\author{Pedram Hassanzadeh$^{1,2}$}\thanks{pedram@rice.edu}
\author{Saleh Nabi$^3$}
\author{Piyush Grover$^3$}

\affiliation{$^1$Department of Mechanical Engineering, Rice University, Houston, TX 77005, USA \\
$^2$Department of Earth, Environmental, and Planetary Sciences, Rice University, Houston, TX 77005, USA \\
$^3$Mitsubishi Electric Research Labs, Cambridge, MA 02139, USA}


\date{\today}

\begin{abstract}

A One-Dimensional (1D) Reduced-Order Model (ROM) has been developed for a 3D Rayleigh-B\'enard convection system in the turbulent regime with Rayleigh number $\mathrm{Ra}=10^6$. The state vector of the 1D ROM is horizontally averaged temperature. Using the Green's Function (GRF) method, which involves applying many localized, weak forcings to the system one at a time and calculating the responses using long-time averaged Direct Numerical Simulations (DNS), the system's Linear Response Function (LRF) has been computed. Another matrix, called the Eddy Flux Matrix (EFM), that relates changes in the divergence of vertical eddy heat fluxes to changes in the state vector, has also been calculated. Using various tests, it is shown that the LRF and EFM can accurately predict the time-mean responses of temperature and eddy heat flux to external forcings, and that the LRF can well predict the forcing needed to change the mean flow in a specified way (inverse problem). The non-normality of the LRF is discussed and its eigen/singular vectors are compared with the leading Proper Orthogonal Decomposition (POD) modes of the DNS data. Furthermore, it is shown that if the LRF and EFM are simply scaled by the square-root of Rayleigh number, they perform equally well for flows at other $\mathrm{Ra}$, at least in the investigated range of $ 5 \times 10^5 \le \mathrm{Ra} \le 1.25 \times 10^6$. The GRF method can be applied to develop 1D or 3D ROMs for any turbulent flow, and the calculated LRF and EFM can help with better analyzing and controlling the nonlinear system.

\end{abstract}


\maketitle


\section{\textrm{I}. Introduction \label{section:Intro}}

Buoyancy-driven turbulence plays a key role in various geophysical and environmental flows such as atmospheric and oceanic circulations as well as engineering systems such as wind farms and Heating, Ventilation, and Air Conditioning (HVAC) technologies. As a result, understanding, predicting, controlling, and optimizing buoyancy-driven turbulence has been of significant interest to the fluid dynamics and climate science communities. Given that Direct Numerical Simulation (DNS) or Large Eddy Simulation (LES) of the full-dimensional Navier-Stokes equations can become computationally prohibitive for fully turbulent flows, which is the relevant regime in most of the aforementioned problems, a considerable attention has been drawn recently to developing Reduced-Order Models (ROMs) for these systems \cite{Liakopoulos1997, Podvin2001, majda2005information, Cuba2011, podvin2012proper, San2015, Annoni2016, Hassanzadeh2016b,Kramer2017}. 

ROMs are low-dimensional models with low computational complexity that retain the necessary dynamics of the turbulent flow, and can be as simple as a system of nonlinear Ordinary Differential Equations (ODEs), or even simpler, linear ODEs, e.g., 
\begin{equation}
	 \dot{\boldsymbol{x}}(t) = \mrm{L} \boldsymbol{x}(t) + \boldsymbol{f}(t) \, . \label{eqn:Intro3}
\end{equation}
where $\boldsymbol{x}$ is the state vector, $\mrm{L}$ is the system's evolution operator or Linear Response Function (LRF), and $\boldsymbol{f}(t)$ represents external forcings (actuations) and/or stochastic parameterization of some unresolved physical processes \citep{holmes2012turbulence,noack2011reduced,palmer1999nonlinear,penland2003stochastic}. This ROM (Eqn.~(\ref{eqn:Intro3})) can be used, for example, to determine the time-mean response of the system to a forcing as $\tm{\boldsymbol{x}} = -\mrm{L}^{-1} \tm{\boldsymbol{f}}$, where $\tm{ \, }$ denotes the long-time average, or to find the forcing required to produce a particular response as $\boldsymbol{f} = -\mrm{L} \tm{\boldsymbol{x}}$ (inverse problem), which can be used for flow control. Furthermore, the spectral properties of $\mrm{L}$ provide information on the dynamics of the system (the limitations and underlying assumptions of Eqn.~(\ref{eqn:Intro3}) are discussed in section~III). 
        
In the fluid dynamics community, the most common model reduction approach is to identify energetically dominant modes, obtained as top eigenvectors from some variant of Proper Orthogonal Decomposition (POD) on the time series, and project the governing equations onto the subspace spanned by these modes \citep{berkooz1993proper,rowley2006model,holmes2012turbulence}. The POD-based methods have been used to study various problems such as wall-bounded shear flows \cite{Aubry1988, Berkooz1993, Moehlis2002}, cavity-driven flows \cite{Cazemier1998, Arbabi2017}, and flows past a cylinder \citep{Ma2002, Noack2006, Rowley2009} to name a few. Several studies have employed POD to develop ROMs for buoyancy-driven flows such as the Rayleigh-B\'enard (RB) convection system \citep{Sirovich1990, Park1990, Deane1991, Sirovich1991, Cuba2011, podvin2012proper}, convection in laterally heated cavities \cite{Gunes1997, Liakopoulos1997, Podvin2001, Benosman2017}, gravity currents \citep{San2015}, and turbulence in wind farms \citep{Andersen2013, Annoni2015, hamilton2016low}. However, because the POD leads to a purely energy-based selection of leading modes, the modes may lack any true dynamical relevance. Furthermore, the truncated (low-energy) modes may still play a crucial role in the dynamics, especially for non-normal systems, where the transient growth can be large \citep{Aubry1988,rowley2017model}. For instance, for examples of buoyancy-driven turbulence, \citet{Cuba2011} and \citet{Benosman2017} showed that owing to the nonlinear interactions between the retained and excluded POD modes, eddy momentum and heat fluxes are not accurately captured, unless some semi-empirical mode-dependent closure models for the viscosity and diffusivity coefficients are employed.   

As an alternative to POD-based methods, calculating $\mrm{L}$ in Eqn.~(\ref{eqn:Intro3}) via the modes of Koopman operator \citep{koopman1931hamiltonian,mezic2005spectral} or their data-driven approximations obtained from Dynamic Mode Decomposition (DMD) \citep{Schmid2008, Rowley2009, Schmid2010, Tu2014, williams2015data, Arbabi2017} has received significant attention and has been applied to a variety of fluid flows, see, e.g., \citet{mezic2013analysis}, \citet{rowley2017model} and references therein. These techniques have also been applied to a number of buoyancy-driven turbulent flows. For instance, \citet{Kramer2017} utilized DMD with sparse sensing to study convection in a laterally heated cavity, \citet{Annoni2016} and \citet{Annoni2017} employed this technique to develop ROMs for two-turbine wind farms in the planetary boundary layer, and \citet{giannakis2018koopman}  conducted Koopman eigenfunction analysis of the 3D flow in a closed cubic turbulent convection cell. While the Koopman/DMD-based methods have produced promising results in these studies, particularly not far from the onset of linear instability, application of these methods to fully turbulent flows, including buoyancy-driven flows, remains a challenge and subject of extensive research. 

Another recently developed framework, known as the resolvent approach, aims to find the perturbations around the turbulent mean flow by knowing the mean profile {\it a priori} and treating the Reynolds stress term in the Navier-Stokes equations as exogenous forcings \cite{McKeon2010, Sharma2013, Moarref2014, Gomez2016}. This unknown forcing is assumed to be connected to the velocity field response via a linear operator called the resolvent. This method, which does not invoke any assumptions with regard to the amplitude of the perturbations, accounts for the nonlinear interaction between different modes through these forcings.

In the climate community, the most common methods for calculating $\mrm{L}$ in Eqn.~(\ref{eqn:Intro3}) are Fluctuation-Dissipation Theorem (FDT) \citep{kubo1966fluctuation,Leith1975,majda2005information} and Linear Inverse Modeling (LIM) \citep{penland1989random,penland2007stochastic}; the latter is closely connected to DMD \citep{Tu2014,Khodkar2018}. Both LIM and FDT are data driven and obtained from the Fokker-Planck equation under certain conditions \citep{penland1989random,majda2005information,Khodkar2018}. While both methods work well when applied to very simple models such as the Lorenz-96 equations, acquiring accurate $\mrm{L}$ for more complex systems such as the quasi-geostrophic equations or Global Circulation Models (GCMs) has been found challenging \cite{gritsun2007climate,Ring2008,cooper2011climate,cooper2013estimation,lutsko2015applying,Hassanzadeh2016b}. 

In a different approach, \citet{Kuang2010} introduced the Green's function (GRF) method, which uses simulations with many weak, localized forcings to construct $\mrm{L}$ (details are presented in section~\ref{section:GRF} IV). He showed that the LRF of a cloud-resolving convection model can be accurately calculated using the GRF method. \citet{Hassanzadeh2016a} extended the GRF method to an idealized GCM and found that the calculated LRF was fairly accurate for the fully turbulent large-scale atmospheric circulation. They further showed that an Eddy Flux Matrix (EFM), $\mrm{E}$, that relates changes in the divergence of turbulent eddy momentum and heat fluxes $\boldsymbol{q}$ to a change in the state $\boldsymbol{x}$ via $\boldsymbol{q} = \mrm{E} \, \boldsymbol{x}$ can be accurately computed as a bi-product of calculating $\mrm{L}$ using the GRF method. In a second study, \citet{Hassanzadeh2016b} used this accurate $\mrm{L}$ to identify the source of inaccuracy in the LRF obtained using FDT as a combination of the GCM operator's non-normality and truncation of the time series to a limited number of POD modes. These accurate LRF and EFM have been also applied to study several aspects of atmospheric circulation in the tropics \citep{kuang2012weakly,herman2013linear} and extratropics \citep{hassanzadeh2015blocking,ma2017quantifying,hassanzadeh2018quantifying}.

Given the success of the GRF method in calculating accurate $\mrm{L}$ and $\mrm{E}$ for fully turbulent atmospheric flows and improving the understanding of the data-driven methods (as mentioned above), it is worthwhile to introduce and examine the GRF method in the context of a canonical fluid dynamics problem that is of broader interest. This is the main purpose of the current study. We also extend the work of \citet{Kuang2010}  and \citet{Hassanzadeh2016a} by showing that, at least for the problem studied here, the LRF and EFM, calculated at a given parameter, can be simply scaled and applied to a wider parameter regime.

\begin{figure}
  \centerline{\includegraphics[width=0.70\textwidth]{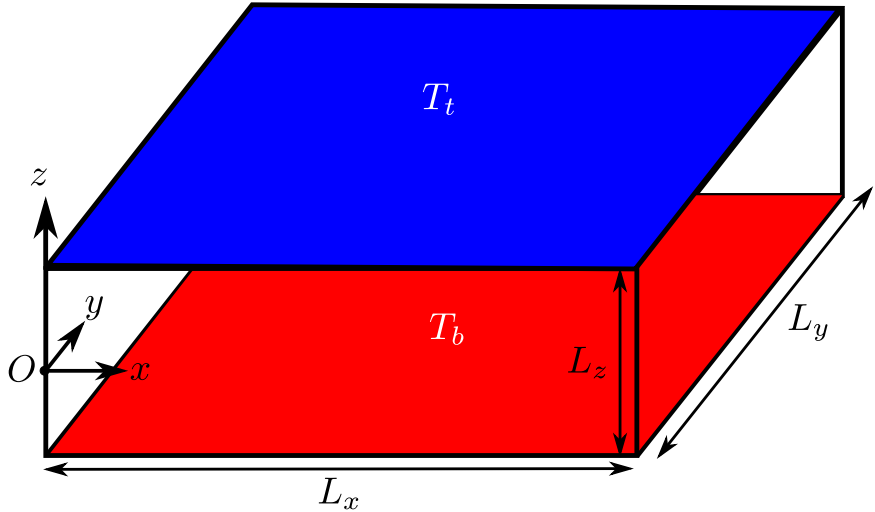}}
  \caption{Schematic of the 3D Rayleigh-B\'{e}nard (RB) convection system. Temperature at the top and bottom walls is, respectively, maintained at constant values of $T_t$ and $T_b$, where $\Delta T=T_b - T_t > 0$. The velocity boundary condition at the walls is no-slip, $\boldsymbol{u}=0$. The horizontal directions ($x$ and $y$) are periodic. In this study, $\mathrm{Pr}=0.707$ and $ 5 \times 10^5 \le \mathrm{Ra} \le 1.25 \times 10^6$.}
\label{fig:Config}
\end{figure}

We have applied the GRF method to a 3D RB convection system (Figure~\ref{fig:Config}) at the Rayleigh number of $\mathrm{Ra}=10^6$, where the flow is far from the onset of linear instability and fully turbulent. The RB convection system is a fitting prototype for buoyancy-driven flows and has been widely used to understand the turbulence physics and to develop techniques for analyzing turbulent systems \citep{ahlers2009heat,doering1996variational,hassanzadeh2014wall,farhat2017continuous,wen2015structure,wen2018reduced}. Focusing on a 1D ROM for the 3D turbulent flow, we have calculated $\mrm{L}$ and $\mrm{E}$ for horizontally averaged temperature and divergence of vertical eddy heat flux at $\mathrm{Ra}=10^6$. Using several tests, we demonstrate that the calculated $\mrm{L}$ and $\mrm{E}$ can predict the response of the system to external forcings accurately. Furthermore, $\mrm{L}$ can calculate the forcing needed to achieve a specified mean flow. While $\mrm{L}$ and $\mrm{E}$ are obtained for $\mathrm{Ra}=10^6$, we show that with a scaling factor that is simply proportional to $\sqrt{\mathrm{Ra}}$, these $\mrm{L}$ and $\mrm{E}$ work accurately at least for $5\times 10^5 \le \mathrm{Ra} \le 1.25 \times 10^6$ as well. 

The structure of this paper is as follows. The mathematical formulation of the RB system and the pseudospectral solver used to conduct DNS are described in section~\ref{section:DNS} II. The 1D ROM is derived in section~\ref{section:Theory} III. The GRF method is presented in section~\ref{section:GRF} IV in detail. The accuracy of $\mrm{L}$ and $\mrm{E}$ for $\mathrm{Ra} = 10^{6}$ and for $5\times 10^5 \le \mathrm{Ra} \le 1.25 \times 10^6$ are discussed in sections~\ref{section:Results} V and \ref{section:Scaling} VI, respectively. The spectral properties of the 1D ROM are investigated in section \ref{section:Spectral} VII. Section~\ref{section:Conclusion} VIII concludes the paper with a brief summary of the present investigation and the outlook for future work.

\color{black}

\section{\textrm{II}. The Boussinesq Equations and Numerical Solver \label{section:DNS}}

We model the turbulent RB convection system using the 3D Boussinesq equations. We non-dimensionalize length with the domain height $L_z$, temperature with $\Delta T = T_b - T_t$, and time with diffusive time scale $\tau_{diff}=L^2_z/\kappa$ where $\kappa$ is the thermal diffusivity, to arrive at the following dimensionless equations
\begin{eqnarray}
\bnabla^* \cdot \boldsymbol{u}^* &=& 0 \label{eq:div} \, , \\
\frac{\partial \boldsymbol{u}^*}{\partial t^*}+(\boldsymbol{u}^* \cdot \bnabla^*)  \boldsymbol{u}^*  &=& - \bnabla^* p^* + \mathrm{Pr} \, \nabla^{*2} \boldsymbol{u}^* + \mathrm{Ra \, Pr} \,(T^*-T^*_{cond}) \hat{\mathbf{e}}_z  \label{eq:mom} \, , \\
\frac{\partial {T}^*}{\partial t^*}+(\boldsymbol{u}^* \cdot \bnabla^*) {T}^*  &=&  \nabla^{*2} {T}^*  \, . \label{eq:temp}
\end{eqnarray}
Here, $\boldsymbol{u}^* = (u^*,v^*,w^*)$ represents the 3D velocity field, $T^*$ shows the temperature, and $T^*_{cond} = 1/2-z^*$ is the conduction temperature profile. Superscript $*$ denotes dimensionless variables and operators hereafter. It should be noted that while $\tau_{diff}$ is used here following convention, we employ the dynamically more relevant advective time scale $\tau_{adv}= \sqrt{L_z /(g \alpha \Delta T)}$ to non-dimensionalize time and vertical velocity when presenting the results ($\tau_{diff}/\tau_{adv}=\sqrt{\mathrm{Ra}\mathrm{Pr}} \approx 840$).

The Rayleigh and Prandtl numbers are defined as
\begin{eqnarray}
\mathrm{Ra} &=& \frac{g \alpha \Delta T L^3_z}{\nu \kappa}  \, , \\
\mathrm{Pr} &=& \frac{\nu}{\kappa} \, ,
\end{eqnarray}
where $g$ represents gravitational acceleration, $\alpha$ and $\nu$ indicate the thermal expansion coefficient, and the kinematic viscosity of the fluid, respectively. The boundary conditions are periodic in the horizontal ($x$ and $y$) directions and fixed temperature and no-slip at the top and bottom walls, i.e.,
\begin{eqnarray}
\boldsymbol{u}^*(x^*,y^*,z^*=\pm1/2,t^*) = 0 \, .
\end{eqnarray}

In this study we use a fixed $\mathrm{Pr}=0.707$ (air), and develop the LRF and EFM for $\mathrm{Ra}=10^6$, which is $\sim 585$ times larger than the critical Rayleigh number for linear instability in this RB setup \citep{drazin2004hydrodynamic}. The flow is fully turbulent at this $\mathrm{Ra}$ (see below). A number of additional tests at a range $ 5 \times 10^5 \le \mathrm{Ra} \le 1.25 \times 10^6$ are also conducted and discussed in section~\ref{section:Scaling} VI.

DNS of Eqns.~(\ref{eq:div})--(\ref{eq:temp}) is carried out using a pseudo-spectral Fourier-Fourier-Chebyshev solver that is based on the code described in \citet{barranco20063d}. Briefly, the solver uses the second-order Adams-Bashforth and Crank-Nicolson schemes for the time integration of the nonlinear and viscous terms, respectively. The no-slip and fixed temperature boundary conditions are enforced following \citet{marcus1984simulation}. Variants of this solver has been used in the past to study geophysical and astrophysical turbulence \citep{barranco2005three, hassanzadeh2012universal,hassanzadeh2013baroclinic,marcus2013three,marcus2015zombie}. The computational domain is $L^*_x \times L^*_y \times L^*_z=\pi \times \pi \times 1$ and the numerical resolution is $128 \times 128 \times 129$. 

\begin{figure}
  \centerline{\includegraphics[width=0.9\textwidth]{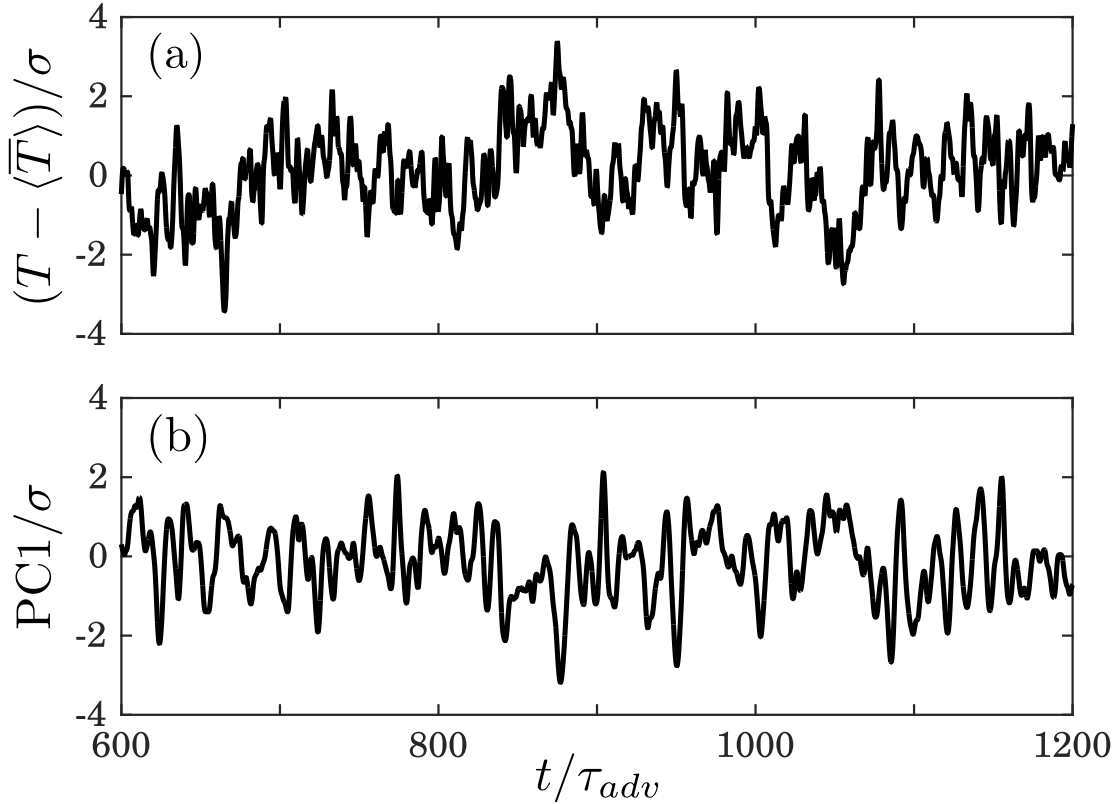}}
  \caption{Time series from DNS at $\mathrm{Ra}=10^6$ for (a) anomalous temperature ($\overline{T}-\langle \overline{T} \rangle$) at $z^*_0 = -0.18$, and (b) the principal component of the leading POD (PC1). Both time series are normalized with their standard deviation ($\sigma$) so that they have the same variability. Time is scaled by the advective timescale $\tau_{adv}$.}
\label{fig:Timeseries}
\end{figure}

\begin{figure}
  \centerline{\includegraphics[width=1.\textwidth]{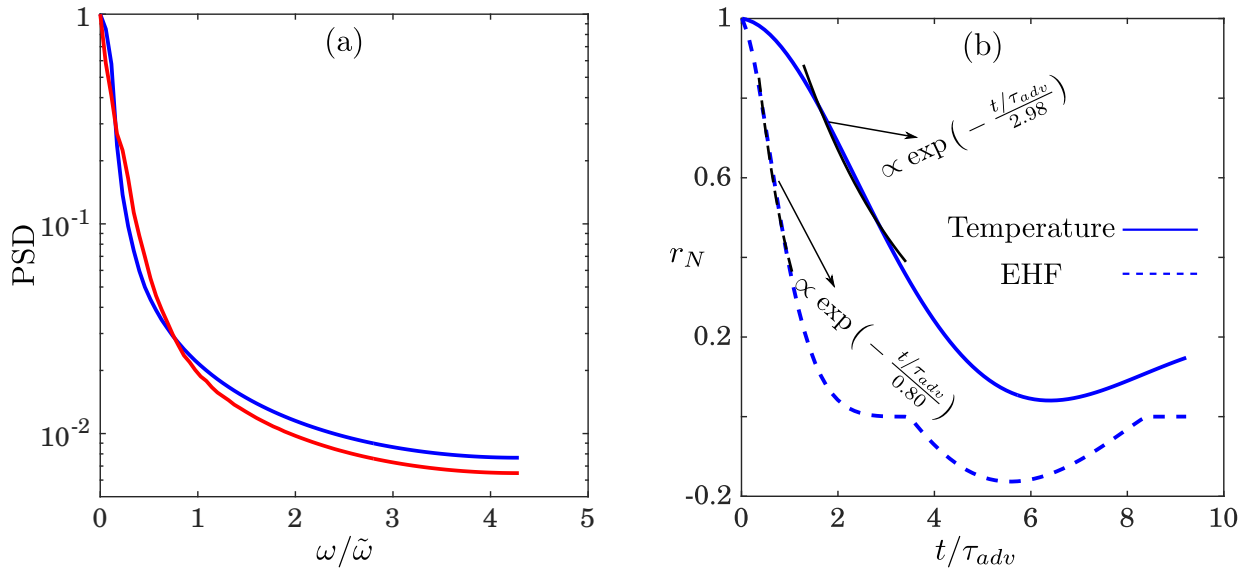}}
  \caption{(a) The Power Spectral Densities (PSDs) of the time series shown in Figure \ref{fig:Timeseries}. The PSDs are calculated by dividing the entire data consisting of $110000$ samples ($\sim 12820\tau_{adv}$) into 500 windows with the same length ($\sim 26\tau_{adv}$), and carrying out fast Fourier transform (FFT) for each window. The results of all windows are then averaged to obtain the plotted PSDs.  Frequency $\omega$ is normalized by the frequency of advective time scale $\tilde{\omega} = 2\pi/\tau_{adv}$. (b) Autocorrelation $r_N$ of anomalous temperatures (solid line) and Eddy Heat Flux, EHF (dashed line), shown as a function of time scaled by $\tau_{adv}$. The black lines show exponential fits.}
\label{fig:AC}
\end{figure}

\begin{figure}
  \centerline{\includegraphics[width=1.\textwidth]{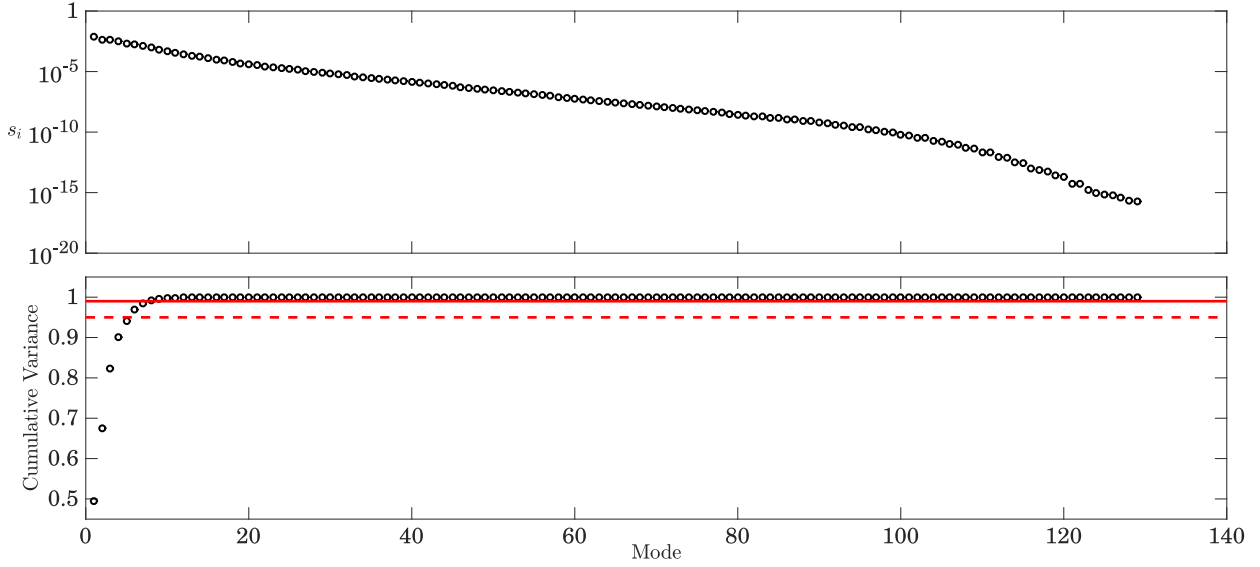}}
  \caption{(a) Singular values $s_i$ of different POD modes obtained by conducting singular value decomposition on the longest available dataset for horizontally averaged anomalous temperature. (b) Cumulative variance explained by the POD modes. The horizontal dashed and solid red lines mark $0.95$ and $0.99$, respectively.}
\label{fig:POD_spectrum}
\end{figure}

For the DNS at $\mathrm{Ra}=10^6$, Fig. \ref{fig:Timeseries} exhibits the time series for the anomalous temperature ($\overline{T}-\langle \overline{T} \rangle$) at $z^*_0 = -0.18$, and the principal component of the leading POD (PC1) obtained via the singular value decomposition of the anomalous temperature. Overbar denotes the spatially averaged variables over the entire $x-y$ plane. These time series illustrate the chaotic nature of the flow, as they show fast oscillations around the mean, and peaks that are a few times larger/smaller than the standard deviation. \color{black} Figure \ref{fig:AC}a demonstrates the power spectra of these two time series, showing that their spectra are monotonically decaying (red spectrum), and do not show any periodic or quasi-periodic behavior, which indicates that the flow is in the fully turbulent regime. To further demonstrate this point, the singular values $s_i$ of different POD modes, and the fraction of variance accumulated up to each POD mode, are shown in Figs. \ref{fig:POD_spectrum}a and b, respectively. As can be seen, no sudden drop in the values of singular numbers occurs, indicating the high-dimensionality of the system, even when the flow is horizontally averaged.

\color{black}

\section{\textrm{III}. 1D ROM for 3D Rayleigh-B\'enard Turbulence \label{section:Theory}}

In the following, we proceed to derive the mathematical formulation of a 1D ROM in the form of Eqn.~(\ref{eqn:Intro3}) for the 3D RB convection system by first averaging all flow properties and equations of motion in the horizontal ($x$ and $y$) directions. The horizontally averaged nonlinear Boussinesq equations can be written as
\begin{equation}
	 \dot{\obf{X}} = F(\obf{X})    \, , \label{eqn:Theory1}
\end{equation}
where $F$ is a nonlinear functional of the state vector $\obf{X}$, which is a set of horizontally averaged variables describing the system. Suppose the state vector evolves from $\tm{\obf{X}}$ at time $t$ to $\tm{\obf{X}} + \obf{x}$ at time $t + \delta t$, in response to an external forcing such as $\obf{f}$. Then Eqn.~(\ref{eqn:Theory1}) yields
\begin{equation}
	 \dot{\obf{x}} = F \left(\tm{\obf{X}} + \obf{x}\right) + \obf{f}   \, . \label{eqn:Theory2}
\end{equation}
If $\obf{x}$ is small, a Taylor expansion of Eqn.~(\ref{eqn:Theory2}) gives 
\begin{equation}
	 \dot{\obf{x}} = \frac{d F}{d \obf{X}}\Big|_{\tm{\obf{X}}} \obf{x} + \obf{f} = \mrm{L} \obf{x} + \obf{f}    \, , \label{eqn:Theory3}
\end{equation}
where the higher order terms (in $\obf{x}$) are neglected (note that $F \left(\tm{\obf{X}}\right) = 0$). Eqn.~(\ref{eqn:Theory3}) shows that the LRF, $\mrm{L}$, is the Jacobian of the nonlinear operator $F$ evaluated at mean state $\tm{\obf{X}}$. To derive this ROM, we do not ignore the eddy-feedback, as we would if we had ignored the nonlinear terms in Eqn.~(\ref{eqn:Theory1}), but in the same fashion as \citet{Ring2008} and \citet{Hassanzadeh2016a}, we assume that a function relating the eddy fluxes and state vector has been linearized and included in $\mrm{L}$ (see below for further discussion). Because we do not know this function, we cannot calculate Eqn.~(\ref{eqn:Theory3}) directly from Eqns.~(\ref{eq:div})-(\ref{eq:temp}). 

It is also instructive to formulate the 1D ROM more explicitly from Eqn.~(\ref{eq:temp}), which combined with Eqn.~(\ref{eq:div}), can be rewritten, in dimensional form, as
\begin{equation}
	 \frac{\partial T}{\partial t} +  \bnabla_\perp  \cdot \left(\boldsymbol{u}_\perp T \right) + \frac{\partial \left(wT\right)}{\partial z} = \kappa \nabla_\perp^2 T + \kappa \frac{\partial^2 T}{\partial z^2}   \, , \label{eqn:Theory4}
\end{equation}
where $\boldsymbol{u}_\perp = u \mathbf{\hat{e}}_x + v \mathbf{\hat{e}}_y$, and $\bnabla_\perp$ and $\nabla_\perp^2$ act only on the $x$ and $y$ directions. Averaging over the $x$ and $y$ directions, and given the periodic boundaries, we find
\begin{equation}
	 \frac{\partial \overline{T}}{\partial t} + \frac{\partial \overline{Tw}}{\partial z} = \kappa \frac{\partial^2 \overline{T}}{\partial z^2}    \, . \label{eqn:Theory5}
\end{equation}
We can decompose $T$ and $w$ into horizontally averaged and around-the-mean perturbation components as $T(x, y, z, t) = \overline{T}(z, t) + T'(x, y, z, t)$ and $w(x, y, z, t) = \overline{w}(z, t) + w'(x, y, z, t)$. Note that $\overline{w}=0$ from continuity. Eqn. (\ref{eqn:Theory5}) can thus be rewritten as
\begin{equation}
	 \frac{\partial \overline{T}}{\partial t} + \frac{\partial \left[ \overline{\left(\overline{T} + T'\right) w'} \right]}{\partial z} = \kappa \frac{\partial^2 \overline{T}}{\partial z^2} \, , \label{eqn:Theory6}
\end{equation}
which further simplifies to
\begin{equation}
	 \frac{\partial \overline{T}}{\partial t} + \frac{\partial \left(\overline{T'w'}\right)}{\partial z} = \kappa \frac{\partial^2 \overline{T}}{\partial z^2}    \, . \label{eqn:Theory7}
\end{equation}
The long-time averaging of Eqn.~(\ref{eqn:Theory7}) leads to
\begin{equation}
	 \left \langle \frac{\partial (\overline{T'w'})}{\partial z}\right \rangle = \kappa \left \langle \frac{\partial^2 \overline{T}}{\partial z^2} \right \rangle    \, . \label{eqn:Theory8}
\end{equation}

Suppose the system evolves from $\tm{\overline{T}}(z)$ to $\tm{\overline{T}}(z) + \ \overline{\theta}(z, t)$ in response to the external forcing $\overline{f}(z, t)$. Eqns.~(\ref{eqn:Theory7}) and (\ref{eqn:Theory8}) then show that the state-vector response $\overline{\theta}$, which represents the horizontal-average of temperature departure from that of the unforced time-mean flow, is governed by
\begin{equation}
	 \frac{\partial \overline{\theta}}{\partial t} + \mrm{E} \overline{\theta} = \kappa \frac{\partial^2 \overline{\theta}}{\partial z^2} + \overline{f}    \, . \label{eqn:Theory9}
\end{equation}
The $\mrm{E} \overline{\theta}$ term represents the change in the divergence of vertical heat flux (second term on the left-hand side of Eqn.~(\ref{eqn:Theory7})) caused by a change in the state $\overline{\theta}$. We emphasize that we do not know the EFM, $\mrm{E}$, and we are not going to make any assumptions about its properties, but we highlight that the representation of the eddy heat flux change via $\mrm{E} \overline{\theta}$ involves two key assumptions:
\begin{enumerate}
    \item The change in the divergence of vertical eddy heat flux, which we denote as $\partial\left(\overline{\theta'w'}\right)/\partial z$ hereafter, can be fully described by  $\overline{\theta}$. This is partly justified if eddies equilibrate rapidly with the new state $\tm{\overline{T}} + \ \overline{\theta}$, which can be evaluated by comparing the auto-correlation timescales of $\overline{\theta}$ and $\partial \left( \overline{\theta'w'} \right) /\partial z$ \citep{Ring2008,Hassanzadeh2016a}. Figure~\ref{fig:AC}b shows the autocorrelation $r_N$ of the times series obtained from projecting $\overline{\theta}$ and $\partial \left( \overline{\theta'w'} \right) /\partial z$ onto the leading POD of $\overline{\theta}$ following \citet{ma2017quantifying}. The results show that the $e$-folding decorrelation timescale of eddies is $\sim 3.7$ times smaller than that of $\overline{\theta}$, suggesting that the eddies decorrelate quickly and equilibrate with the new state, e.g., after $3 \tau_{adv}$, the ratio of $r_N$ of eddies and $\overline{\theta}$ is $0.064$

    \item $\partial \left( \overline{\theta'w'} \right) /\partial z$ changes linearly with $\overline{\theta}$. This is a reasonable assumption if $\overline{\theta}$ has small amplitude, and consistent with the assumption under which Eqn. (\ref{eqn:Theory3}) was derived.         
\end{enumerate}

In summary, Eqn.~(\ref{eqn:Theory9}) shows that state vector $\obf{x}=\obf{\theta}$ describes the response of the system and that the 1D ROM is
\begin{equation}
	 \dot{\obf{\theta}} = \mrm{L} \obf{\theta} + \obf{f}    \, , \label{eqn:Theory10}
\end{equation}
where $\mrm{L} = \kappa \mrm{D}^2 - \mrm{E}$. The operator $\mrm{D}^2$ is the second derivative with respect to $z$. We show in the next section that the matrix $\mrm{L}$ (and matrix $\mrm{E}$) in Eqn.~(\ref{eqn:Theory10}) can be accurately calculated for a fully turbulent flow using the GRF method without any need for explicit knowledge or approximation of $\mrm{E}$.  

\color{black}

\section{\textrm{IV}. The Green's Function (GRF) Method \label{section:GRF}}

In order to calculate $\mrm{L}$ and $\mrm{E}$ at $\mathrm{Ra}=10^6$, we follow the procedure described in \cite{Hassanzadeh2016a}. First, we define a set of Gaussian basis functions of the form 
\begin{eqnarray}
B_n(z)=\exp{\left[ -\frac{(z - z_n)^2}{z_w^2} \right]} 
\label{eq:basis}
\end{eqnarray}
where $z_w=L_z/20$, $z_n = \{-1, -0.95, -0.9, -0.85, -0.8, -0.7, ... \,  0  \, ... , 0.7, 0.8, 0.85, 0.9, 0.95, 1\}\times L_z/2$, and $n = 1, 2, ... \, 25$. Simpler choices for $z_n$ such as $z_n = \{-1, -0.9, -0.8, \, ... \, , 0.8, 0.9, 1 \}\times L_z/2$ were initially tried, but it was realized that in order to develop a reasonably accurate ROM, basis functions should be denser near the walls to better resolve the sharp gradients in the boundary layers. We calculate $\mrm{L}$ and $\mrm{E}$ in the space of these 25 basis functions rather than for the entire grid space (129 points) to reduce the computational cost.   

Second, forcings of the form $f_n(z)=a_n \Delta T/\tau_{diff} \times B_n(z)$ are added to the right-hand side of Eqn.~(\ref{eq:temp}) one at a time, and a long DNS is then conducted at $\mathrm{Ra}=10^6$. $\Delta T=T_b - T_t$ is the temperature difference between the bottom and top walls (Fig.~\ref{fig:Config}). $a_n$ varies with $z_n$ and is stronger near the walls. Its value is chosen, after some trial and error, such that it is not too large to violate the linearity assumption in Eqns. (\ref{eqn:Theory3}) and (\ref{eqn:Theory10}), or too small, so that the signal (i.e., $\tm{\overline{\theta}}$) to noise (i.e., standard deviation of $\overline{\theta}$) ratio becomes large. To obtain large signal-to-noise-ratio within the linear regime, we have conducted long DNS that are on average nearly $3200$ times longer than $\tau_{adv}$ after the system reaches quasi-equilibrium. Signal-to-noise-ratio and the degree of nonlinearity are quantified using the criteria defined in \citet{Hassanzadeh2016a}. Based on these criteria, $a_n$ is chosen to be 20 for all cases except the first three near-the-wall basis functions for which $a_n = 40$. 

Hereinafter, we refer to each forced DNS as a ``trial''.  To increase the accuracy of the calculated ROM \citep{Kuang2010,Hassanzadeh2016a}, for each $f_n$, one trial with positive and one trial with negative forcing is conducted, and the time-mean response $\tm{\overline{\theta}}$ is calculated. Half of the difference between $\tm{\overline{\theta}}$ for the positive and negative forcings is used as net response to $f_n$ (denoted as $\tm{\obf{\theta}}_n$). Given the symmetries of Eqns.~(\ref{eq:div})-(\ref{eq:temp}), we have only conducted the trials for the lower half of the system $-1 \le z_n \le 0$ ($n=1, 2, ... \, 13$), and just used $\tm{\obf{\theta}}_n(z)=\tm{\obf{\theta}}_{(26-n)}(-z)$ for $n=14, 15, ... \, 25$. Therefore, a total of 26 DNS are needed.  

Each $\tm{\overline{\theta}}_n$ is projected via least-square linear regression onto the basis function space. The resulting projection coefficients are $\tm{\obf{\theta}}_n$ ($n = 1, 2, ... 25$), each of which is a column vector with the length $25$. $\obf{f}_n$ is also a column vector with the same length, whose elements are all zero, except for its $n^{\mathrm{th}}$ element, which is equal to the amplitude of the forcing $f_n$. We can thus construct the following matrices for the time-mean responses and forcings in the reduced dimension of 25
\begin{equation}
\mrm{R} = \left[
	\begin{array}{cccc}
		\tm{\obf{\theta}}_1  & \tm{\obf{\theta}}_2 & ... & \tm{\obf{\theta}}_{25}     
	\end{array}
\right]  \, , \label{eqn:Theory11}
\end{equation}
\begin{equation}
\mrm{F} = \left[
	\begin{array}{cccc}
		\obf{f}_1  & \obf{f}_2 & ... & \obf{f}_{25}  
	\end{array}
\right] \, . \label{eqn:Theory12}
\end{equation}
The LRF $\mrm{L}$ of the system is then calculated from the long-time averaged Eqn.~(\ref{eqn:Theory10}) as 
\begin{equation}
\mrm{L} = -\mrm{F} \mrm{R}^{-1}  \, . \label{eqn:Theory13}
\end{equation}

The EFM, $\mrm{E}$, is evaluated from the same simulations using a similar procedure. $\tm{T'w'}$ is calculated for each trial and the net response to each $f_n$ (denoted as $\tm{\theta'w'}_n$) is obtained from the positive- and negative-forcing trials. The vertical derivative of the eddy flux responses is then calculated and projected onto the basis functions to obtain
\begin{equation}
\mrm{Q} = \left[
	\begin{array}{cccc}
		\Big \langle \frac{\partial ( \obf{\theta' w'} )}{\partial z}\Big \rangle_1 & \Big \langle \frac{\partial ( \obf{\theta' w'} )}{\partial z}\Big \rangle_2 & ... & \Big \langle \frac{\partial ( \obf{\theta' w'} )}{\partial z}\Big \rangle_{25}
	\end{array}
\right] \, . \label{eqn:Theory14}
\end{equation}
$\mrm{E}$ is then computed as
\begin{equation}
\mrm{E} = \mrm{Q} \mrm{R}^{-1}  \, . \label{eqn:Theory15}
\end{equation}
The accuracy and predictive capabilities of $\mrm{L}$ and $\mrm{E}$ presented here are examined in detail for several test cases in section~\ref{section:Results} V.

\color{black}

\section{\textrm{V}. Comparison of the GRF-based ROM and DNS at $\mathrm{Ra}=10^6$ \label{section:Results}}

In the following section, we assess the accuracy of $\mrm{L}$ and $\mrm{E}$ obtained using the GRF method by examining their capabilities to predict the time-mean response of temperature and vertical eddy heat flux to external forcings. Furthermore, we study the performance of $\mrm{L}$ in calculating the forcing required for the control of time-mean flow. To find the ``true'' responses or to evaluate the accuracy of the calculated forcing, long, forced DNS are conducted. The details of all these test cases (denoted by `C') are presented in Table I. Some of the forcings used in these test cases are localized, e.g., the Gaussian forcing of C1, but most of them are in the form of cosine or sinusoidal functions, which excite the flow along the $z$ direction and can lead to complex responses. For example, cosine forcings are strong at the boundaries and lead to large eddy heat flux responses at the boundary layers, and forcings with high wave numbers  create multiple contiguous stabilized and destabilized regions in the domain.   

\begin{table}
\label{table:Tests}
\centering
\caption{Details of the test cases to be discussed in the present and next sections. $f$ is the external forcing for each test case and $t_{\mrm{DNS}}$ indicates the length of the DNS dataset. For most cases, DNS with $\pm f$ (as well as control) are conducted to increase the accuracy and ensure the linearity of the response, but to reduce the computational cost, for a few cases, indicated by ${}^*$, only positive forcing and control DNS are conducted. Error for temperature responses is defined as $\norm{\tm{\obf{\theta}}_{\mrm{GRF}} - \tm{\obf{\theta}}_{\mrm{DNS}}}_2/\norm{\tm{\obf{\theta}}_{\mrm{DNS}}}_2$. Similar formulation is used to find the error in the divergence of the vertical Eddy Heat Flux (EHF) response.} 
\centering
\begin{tabularx}{0.90\textwidth}{|c|c|c|c|c|c|c|}
  \hline
  Case   &          Figure                &   $\mathrm{Ra}$    &        $f/(\Delta T / \tau_{diff} )$          & $\theta$ error ($\%$) & EHF error ($\%$) & $ t_{DNS}/\tau_{adv} $ \\
  \hline
  C1     & \ref{fig:GRF}a $\&$ b     & $10^6$             & $10 \exp\Big[-\frac{(z^* - 0.2)^2}{0.1^2} \Big]$  &             2.28       &    16.35  & 2307 \\ \hline
  C2     &     Not shown                  & $10^6$        &      $20 \cos(2\pi z^*)$                            &             5.35       &     6.08  & 2496 \\ \hline
  C3     & \ref{fig:GRF}c $\&$ d     & $10^6$             & $10 \sin(2\pi z^*)$                               &             9.32       &     4.84  & 2460 \\ \hline
  C4     & \ref{fig:GRF}e $\&$ f     & $10^6$             & $10 \cos(8 \pi z^*)$                              &             9.45       &    16.82  & 3133 \\ \hline
  C5     & \ref{fig:Inv}a $\&$ b     & $10^6$             & As in Fig. \ref{fig:Force}a                        &             6.07       &     8.28  & 2554 \\ \hline
  C6     & \ref{fig:Inv}c $\&$  d    & $10^6$             & As in Fig. \ref{fig:Force}b                        &             19.41      &    11.05  & 2772 \\ \hline
  C7$^*$ &      Not shown            & $5 \times 10^5$    & $10 \exp\Big[-\frac{(z^* - 0.2)^2}{0.1^2} \Big]$  &             3.49       &    13.44  & 2135 \\ \hline
  C8$^*$ & \ref{fig:Scaling}a $\&$ b & $5 \times 10^5$    & $10 \cos(2\pi z^*)$                               &             4.36       &     7.02  & 2150 \\ \hline
  C9$^*$ & \ref{fig:Scaling}c $\&$ d & $7.5 \times 10^5$  & $ 20 \cos(2\pi z^*)$                                &             4.87       &    8.92   & 2543 \\ \hline
  C10    & \ref{fig:Scaling}e $\&$ f & $1.25 \times 10^6$ & $ 20 \exp\Big[-\frac{(z^* - 0.2)^2}{0.15^2} \Big]$   &             3.91       &    17.22  & 2719 \\ 
  \hline
\end{tabularx}
\end{table}


\begin{figure}
  \centerline{\includegraphics[width=.8\textwidth]{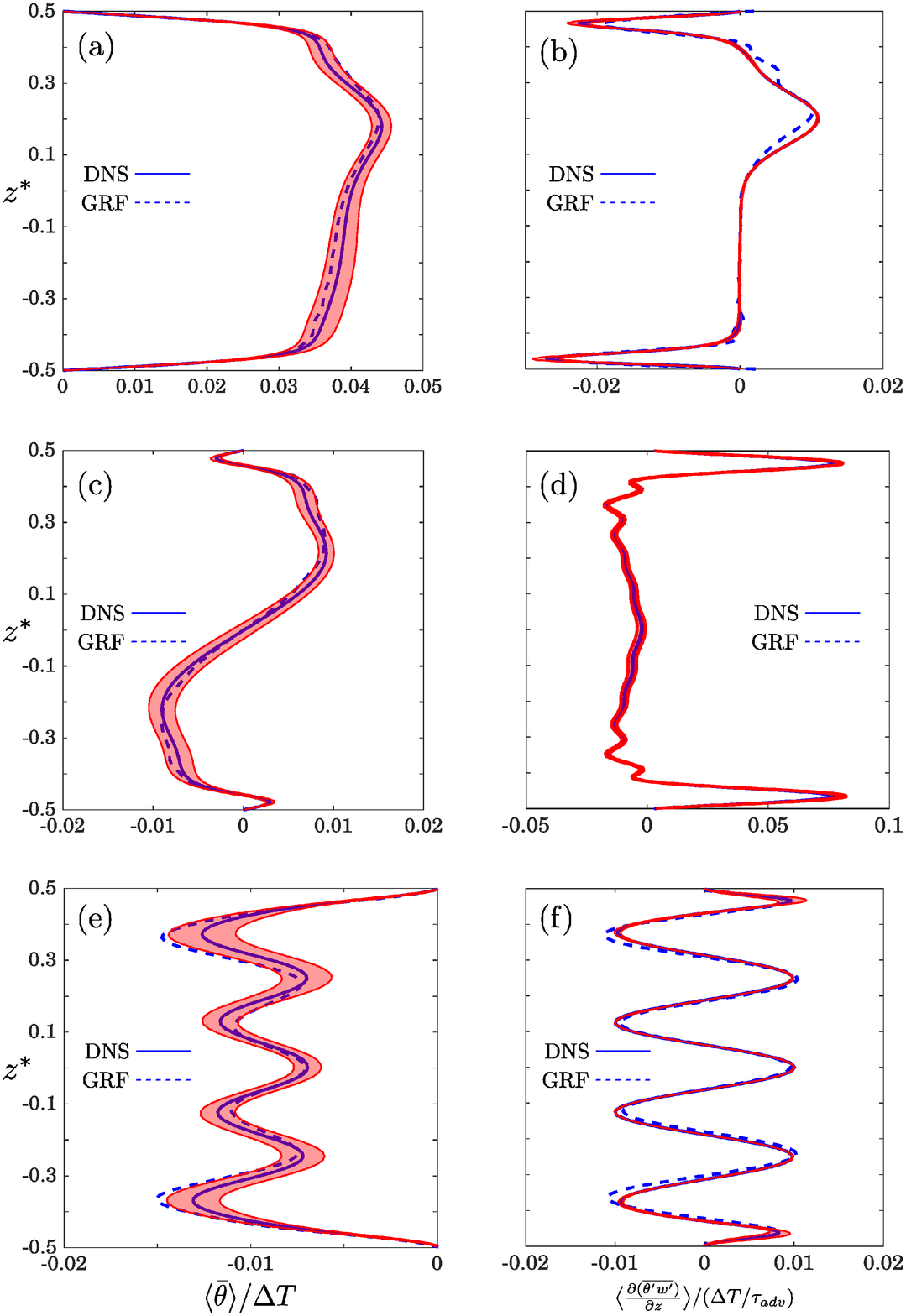}}
  \caption{Time-mean responses of temperature (left column) and eddy heat flux (right column) to forcings (a-b) $f_1$, (c-d) $f_3$, and (e-f) $f_4$ (see Table~I). Solid lines show the ``true'' response obtained from long forced DNS, and dashed lines show the predictions from $\mrm{L}$ and $\mrm{E}$ obtained using the GRF method. The red shading shows the uncertainty in the time-mean responses calculated from the DNS data (see the text for more details). For all cases, $\mathrm{Ra} = 10^6$.}
\label{fig:GRF}
\end{figure}

Figure~\ref{fig:GRF} compares the ROM and DNS results for the time-mean response of horizontally averaged temperature $\tm{\overline{\theta}}$, scaled by $\Delta T$, and eddy heat fluxes $\big \langle \partial (\overline{\theta' w'})/\partial z\big \rangle$ scaled by $\Delta T/\tau_{adv}$, for three different cases (C1, C3, and C4; C2 is not shown for brevity). The red shadings in this figure and other figures demonstrate the uncertainty in the time-mean responses calculated from DNS. To find this uncertainty, each DNS time series is divided into eight segments with equal length, and the standard deviation $\sigma$ of the time-mean of these segments are calculated. The solid blue lines show the mean of these eight segments, while the shading shows $\pm \sigma$. As shown in Fig.~\ref{fig:GRF}, despite the notable complexity of some of the responses such as sharp gradients in the boundary layers and multiple extrema, the pattern and amplitude of the temperature and eddy heat flux responses are well predicted by $\mrm{L}$ and $\mrm{E}$ in all these test cases.    

\begin{figure}
  \centerline{\includegraphics[width=.8\textwidth]{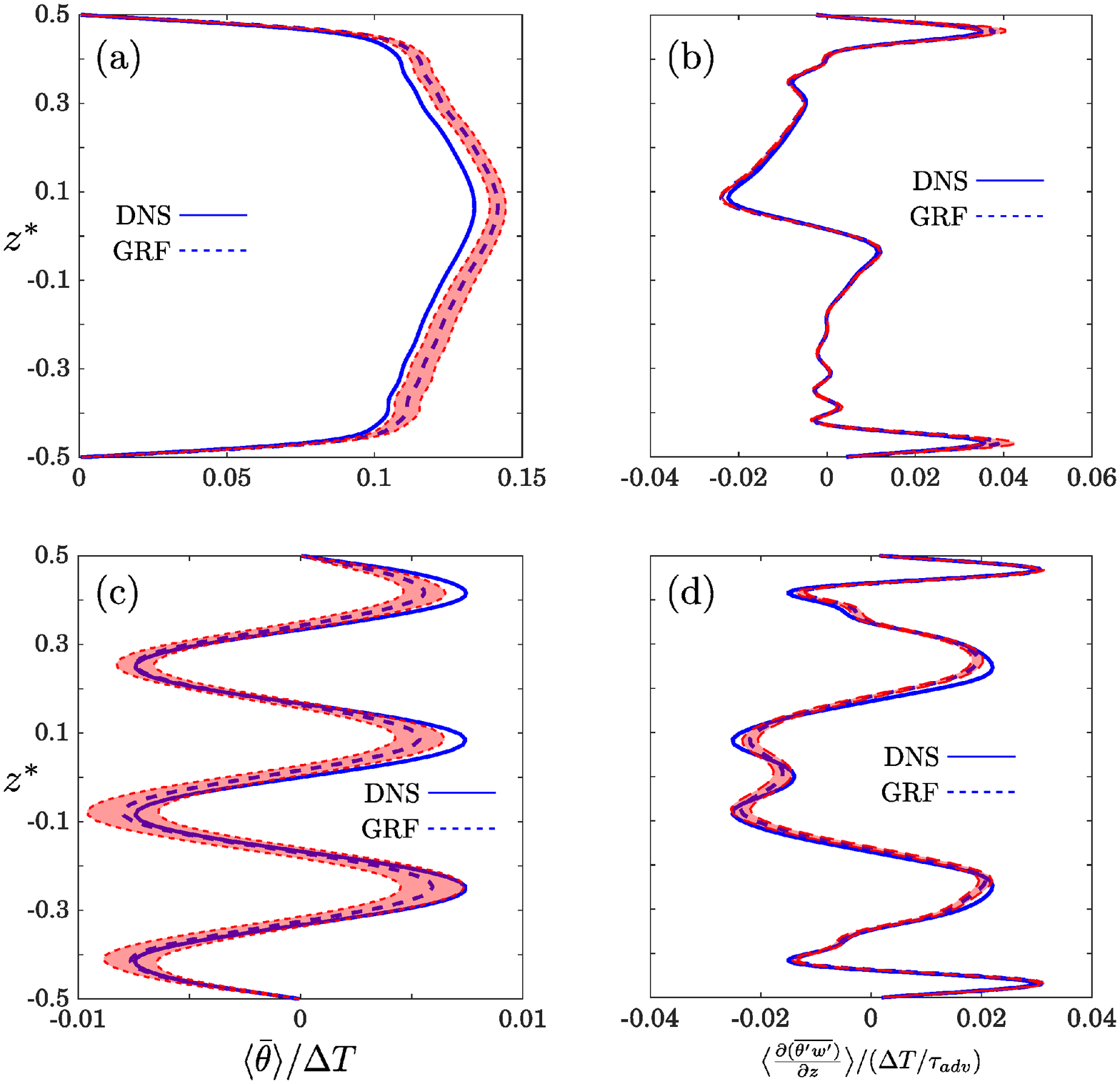}}
  \caption{(a)-(b) C5; (c)-(d) C6. (a) and (c): Solid lines show the target change in the mean-flow $\theta_{\mrm{target}}$ while the dashed lines demonstrate the time-mean responses to forcings $f_5$ (shown in Fig. \ref{fig:Force}a) and $f_6$ (shown in Fig. \ref{fig:Force}b) obtained from long, forced DNS. (c) and (d): changes in vertical eddy heat flux obtained from long, forced DNS (solid lines) or calculated by $\mrm{E}$ (dashed lished). As before, the red shading shows the uncertainty in the time-mean responses calculated from the DNS data. For all cases, $\mathrm{Ra} = 10^6$. More details are in Table I.}
\label{fig:Inv}
\end{figure}

\begin{figure}
  \centerline{\includegraphics[width=1.\textwidth]{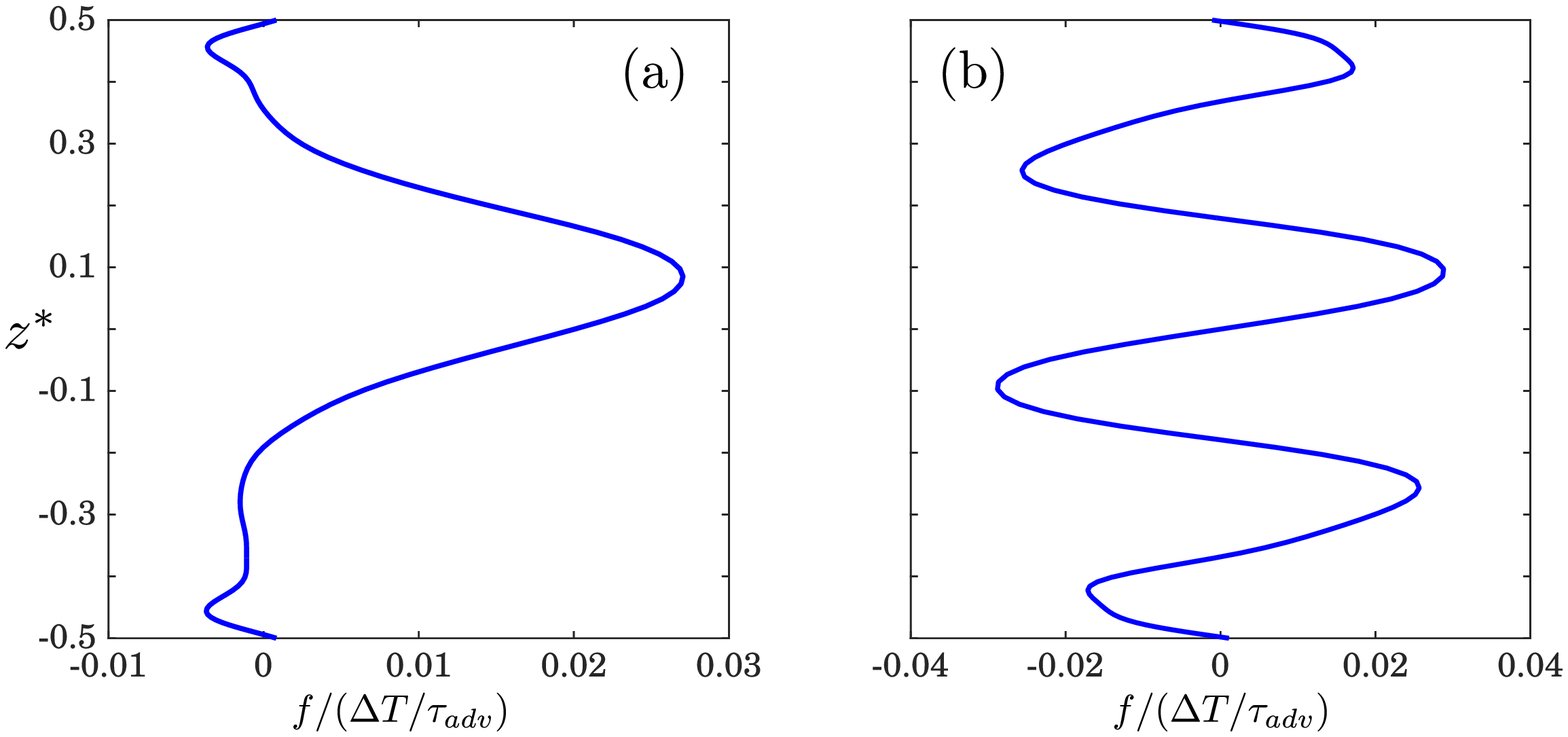}}
  \caption{The forcing that is predicted as $f=-\mrm{L} \theta_{target}$ to lead to the target time-mean responses $\theta_{target}$. (a) $f_5$ ($\theta_{target}$ is shown in Fig.~\ref{fig:Inv}a), and (b) $f_6$ ($\theta_{target}$ is shown in Fig.~\ref{fig:Inv}c).}
\label{fig:Force}
\end{figure}

Knowing $\mrm{L}$ also enables us to find the required forcing to produce a desired change in the time-mean flow. This is of particular interest when flow control is intended. To test the skill of $\mrm{L}$ for such inverse problems, we have chosen two target profiles $\theta_{target}$ shown with solid lines in Fig.~\ref{fig:Inv}a (C5) and Fig.~\ref{fig:Inv}c (C6). The forcings needed to change the mean-flow by $\theta_{target}$ for these two cases are calculated as $f=-\mrm{L} \theta_{target}$ and shown in Fig.~\ref{fig:Force}. The forcing profiles are not trivial, particularly near the walls, even for the simpler $\theta_{target}$ of C5. To evaluate the accuracy of these predicted forcings, forced DNS with $f_5$ and $f_6$ are conducted and the mean-flow changes are shown in Fig.~\ref{fig:Inv}a and c (dashed lines), which match the target well, although the amplitude is larger for C5. The accuracy of $\mrm{E}$ can further be examined using these test cases as shown in Fig.~\ref{fig:Inv}b and d. As before, we find that $\mrm{E}$ can well capture changes in the vertical eddy heat flux even for complex $\big \langle \partial (\overline{\theta' w'})/\partial z\big \rangle$ profiles.

\color{black}

\section{\textrm{VI}. Extending the 1D ROM to other values of $Ra$ \label{section:Scaling}} 

In the previous section, we showed that $\mrm{L}$ and $\mrm{E}$ that are calculated using the GRF method at $\mathrm{Ra}=10^6$ work well in predicting the response or forcing at this value of $\mrm{Ra}$. As will be discussed in section~VIII, the main drawback of the GRF method is that it is computationally expensive, therefore, it is worthwhile to explore how the $\mrm{L}$ and $\mrm{E}$ calculated for one value of $\mathrm{Ra}$ can be used for other $\mrm{Ra}$ numbers. 

\begin{figure}
  \centerline{\includegraphics[width=0.8\textwidth]{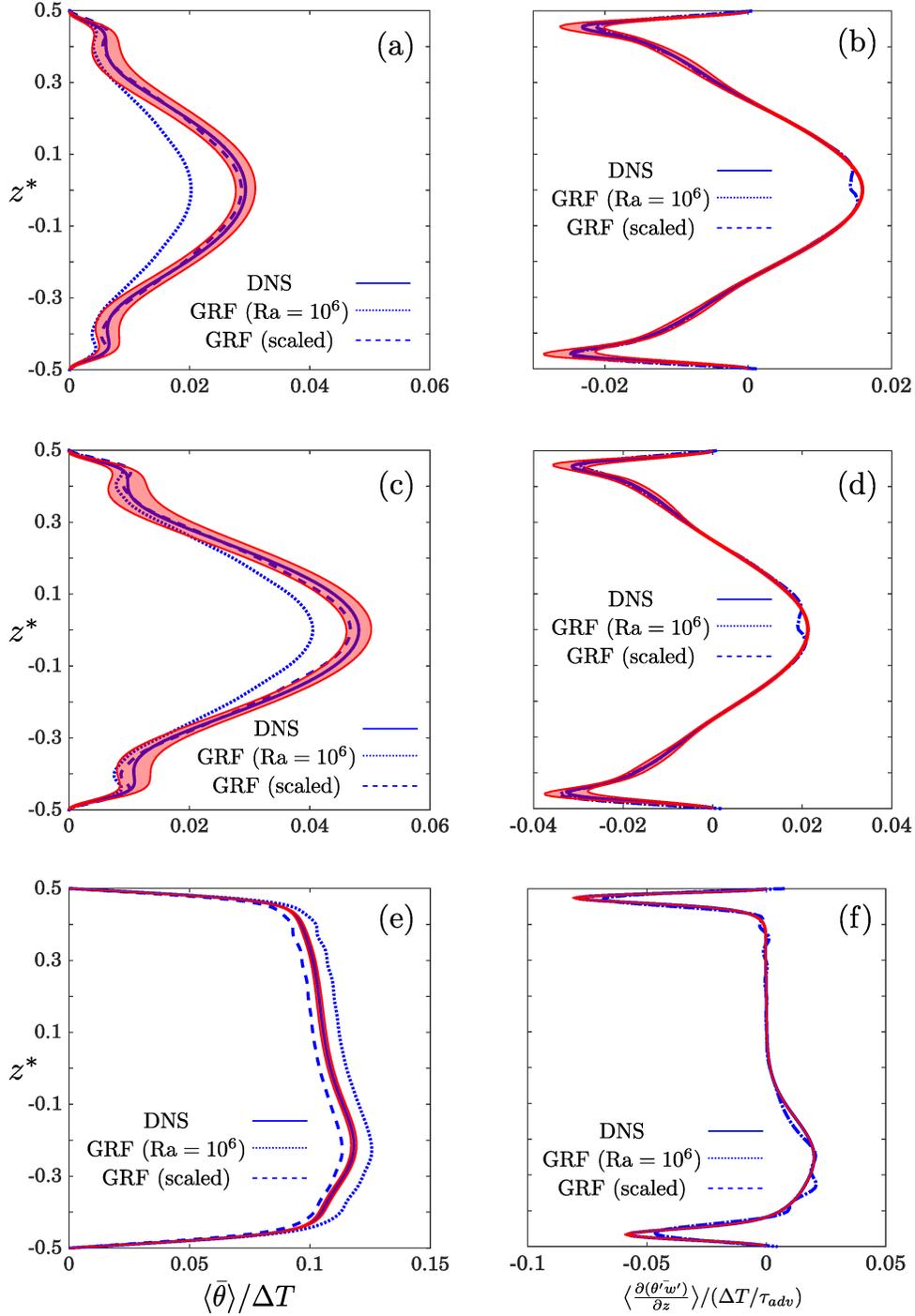}}
  \caption{
  Time-mean responses of temperature (left column) and eddy heat flux (right column) to forcings (a-b) $f_8$ ($\mrm{Ra} = 5 \times 10^5$) (c-d) $f_9$ ($\mrm{Ra} = 7.5 \times 10^5$), and (e-f) $f_{10}$ ($\mrm{Ra} = 1.25 \times 10^6$) (see Table~I). Solid lines show the ``true" response obtained from long forced DNS, dotted lines show the predictions from $\mrm{L}(10^6)$ and $\mrm{E}(10^6)$, and dashed lines show the predictions obtained by the LRF and EFM scaled according to Eqn.~(\ref{eqn:Scaling4}).}
\label{fig:Scaling}
\end{figure}

We have conducted several more forced DNS within the range of $5 \times 10^6 \leq \mathrm{Ra} \leq 1.25 \times 10^6$. Details of some of these simulations are presented in Table I (C7-C10). The solid lines in Fig. \ref{fig:Scaling} show the time-mean responses in temperature and vertical eddy heat flux while the dotted lines show the predictions when the LRF and EFM of $\mathrm{Ra} = 10^6$ are used. For $\tm{\overline{\theta}}$, while the general shapes of the profiles are well captured by $\mrm{L} (10^6)$, the amplitudes are under- or over-estimated, depending on $\mathrm{Ra}$. These results suggest that the eigenvectors of $\mrm{L}$ have remained fairly unchanged for this range of $\mathrm{Ra}$, and that only its eigenvalues have varied. For the eddy heat flux, we find that if the response is calculated as $\mrm{E}(10^6)\mrm{L}^{-1}(10^6)f$, then the prediction is surprisingly accurate (Fig. \ref{fig:Scaling}b, d, and f), indicating that $\mrm{E}\mrm{L}^{-1}$ remains approximately constant for the aforementioned range of $\mrm{Ra}$. We highlight that this hypothesis is not based only on these four observations, but more simulations in the range of $5 \times 10^5 \leq \mathrm{Ra} \leq 1.25 \times 10^6$ confirmed this hypothesis as well. 

Based on the these observations, we postulate that $\mrm{L}(10^6)$ and $\mrm{E}(10^6)$ can be simply scaled to find the LRF and EFM at a new $\mathrm{Ra}$
\begin{eqnarray}
	\mrm{L}(\mathrm{Ra}) = c_f \mrm{L}(10^6) \qquad \, , \qquad \mrm{E}(\mathrm{Ra}) = c_e \mrm{E}(10^6)    \, . \label{eqn:Scaling1} 
\end{eqnarray} 
Furthermore, the fact that $\mrm{E} \mrm{L}^{-1}$ remains nearly constant suggests that the scaling factors are the same: $c_f=c_e$.

To validate Eqn.~(\ref{eqn:Scaling1}), scaling factors are found as 
\begin{eqnarray}
	c_f(\mathrm{Ra}) & = & \frac{\big| \big| \tm{\obf{\theta}}_{\mrm{GRF}} \big| \big|_\infty}{\big| \big| \tm{\obf{\theta}}_{\mrm{DNS}}\big| \big|_\infty}  \, , \\ \label{eqn:Scaling2}
	c_e(\mathrm{Ra}) & = & \frac{\big| \big| \langle \partial (\obf{\theta' w'})/\partial z\big \rangle_{\mrm{GRF}} \big| \big|_\infty}{\big| \big| \langle \partial (\obf{\theta' w'})/\partial z\big \rangle_{\mrm{DNS}}  \big| \big|_\infty} \, , \label{eqn:Scaling3}
\end{eqnarray}
where subscript $\mrm{GRF}$ in the numerators indicates that $\mrm{L}(10^6)$ and $\mrm{E}(10^6)$ are employed, and subscript $\mrm{DNS}$ in the denominators shows results from long, forced DNS at $\mathrm{Ra}$ are used.

\begin{figure}
  \centerline{\includegraphics[width=0.5\textwidth]{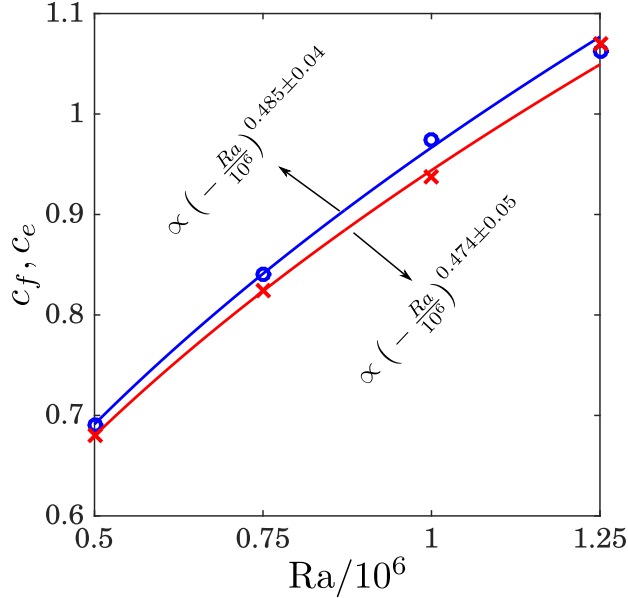}}
  \caption{Scaling factors $c_f$ and $c_e$ demonstrated by blue circles and red crosses, respectively. The solid lines are the power fits to the corresponding discrete data whose functions are shown in the figure.}
\label{fig:Fit}
\end{figure}

Figure \ref{fig:Fit} shows the scaling factors calculated for C2 and C8-C10 along with the power fit to each of them. We find that $c_f \sim c_e$ and that both are approximately $0.5$, suggesting the scaling with $\sqrt{\mathrm{Ra}}$. Therefore we can reasonably approximate the scaled $\mrm{L}$ and $\mrm{E}$ as
\begin{eqnarray}
	\mrm{L}(\mathrm{Ra_2}) = \sqrt{\frac{\mrm{Ra_2}}{\mathrm{Ra_1}}} \, \mrm{L}(\mrm{Ra_1})  \qquad \, , \qquad \mrm{E}(\mathrm{Ra_2}) = \sqrt{\frac{\mrm{Ra_2}}{\mathrm{Ra_1}}} \, \mrm{E}(\mathrm{Ra_1})   \, . \label{eqn:Scaling4} 
\end{eqnarray} 

Dashed lines in Fig. \ref{fig:Scaling} demonstrate the performance of $\mrm{L}$ and $\mrm{E}$ calculated using Eqns. (\ref{eqn:Scaling4}), for three different test cases with $\mathrm{Ra_1}=10^6$. As shown in this figure, predicted responses agree closely with those of the DNS results, which substantiates the validity of  the scaling argument presented earlier for a fairly broad range of Rayleigh numbers. We also highlight that the accuracy of the scaled LRFs and EFMs for a given $\mathrm{Ra}$ is comparable to the accuracy of the $\mrm{L}(10^6)$ and $\mrm{E} (10^6)$. Whether this scaling holds for a larger range of $\mathrm{Ra}$ is computationally expensive to test, and is left for future work.

\color{black}

\section{\textrm{VII}. Spectral properties of the 1D ROM \label{section:Spectral}}

\begin{figure}
  \centerline{\includegraphics[width=.96\textwidth]{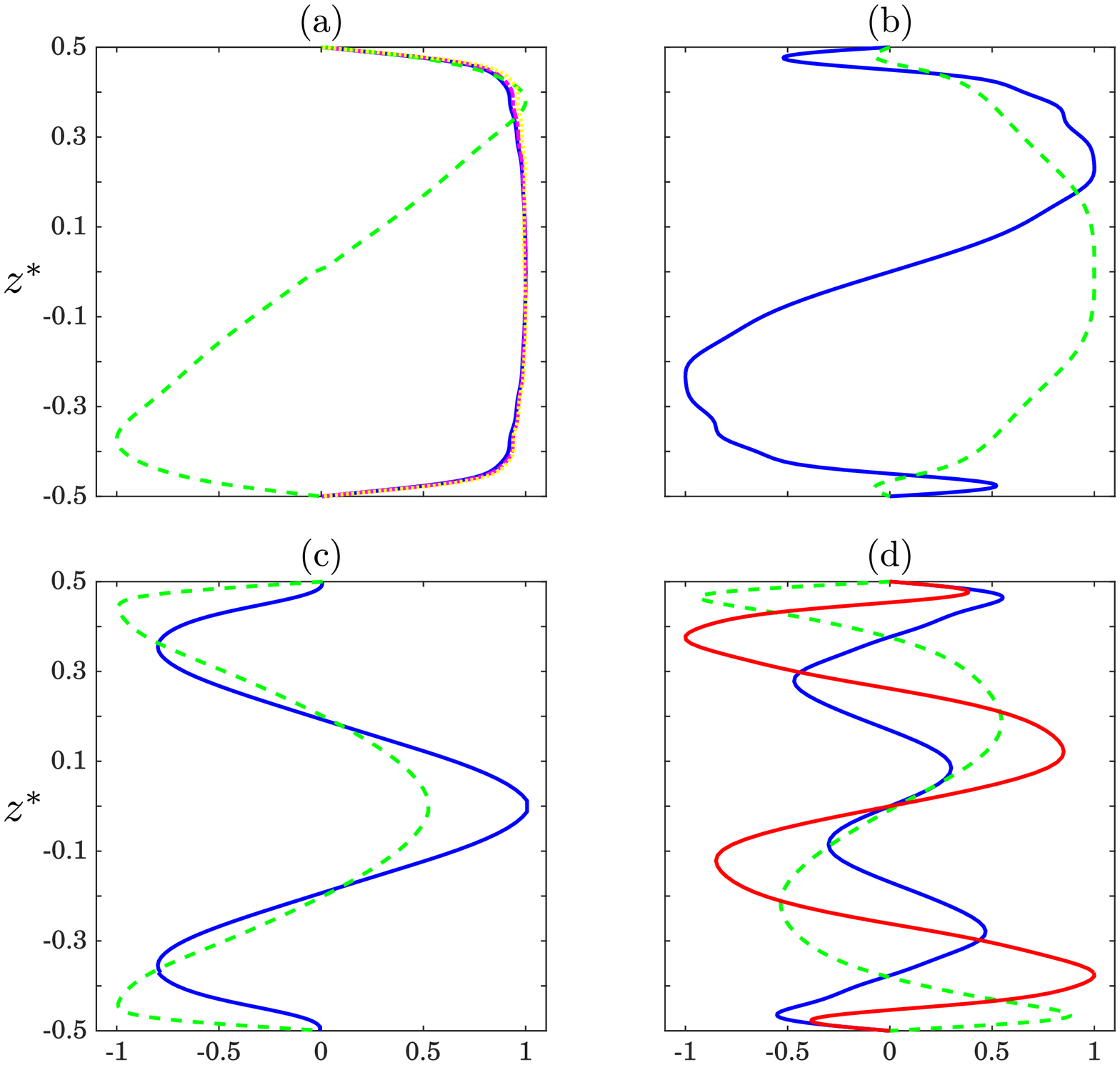}}
  \caption{The first four leading (i.e., slowest-decaying) modes and neutral vector of $\mrm{L}$ and the leading POD modes. Solid blue (red) lines show the real (imaginary) part of $\mrm{L}$'s eigenvectors. The green dashed lines show the leading POD modes (time-mean removed) obtained from the unforced DNS. The first four leading POD modes of unforced DNS explain around $48 \% $, $20\% $, $15\% $, and $7.5 \% $. In panel (a), the magenta dash-dot and yellow dotted line display, respectively, the neutral vector of $\mrm{L}$ and POD1 of the data obtained from the  stochastic ODE (\ref{eqn:Spectral1}); both lines coincide with the eigenvector. The eigenvalues of the shown eigenmodes are (a) $-0.058~1/\tau_{adv}$, (b) $-1.303~1/\tau_{adv}$, (c) $-1.506~1/\tau_{adv}$, and (d) $(-2.726 + 0.504i)~1/\tau_{adv}$ (see Fig.~\ref{fig:EVs}). All these modes are projected onto the grid space and normalized to have the magnitude of one.}
\label{fig:Modes}
\end{figure}

\begin{figure}
  \centerline{\includegraphics[width=.96\textwidth]{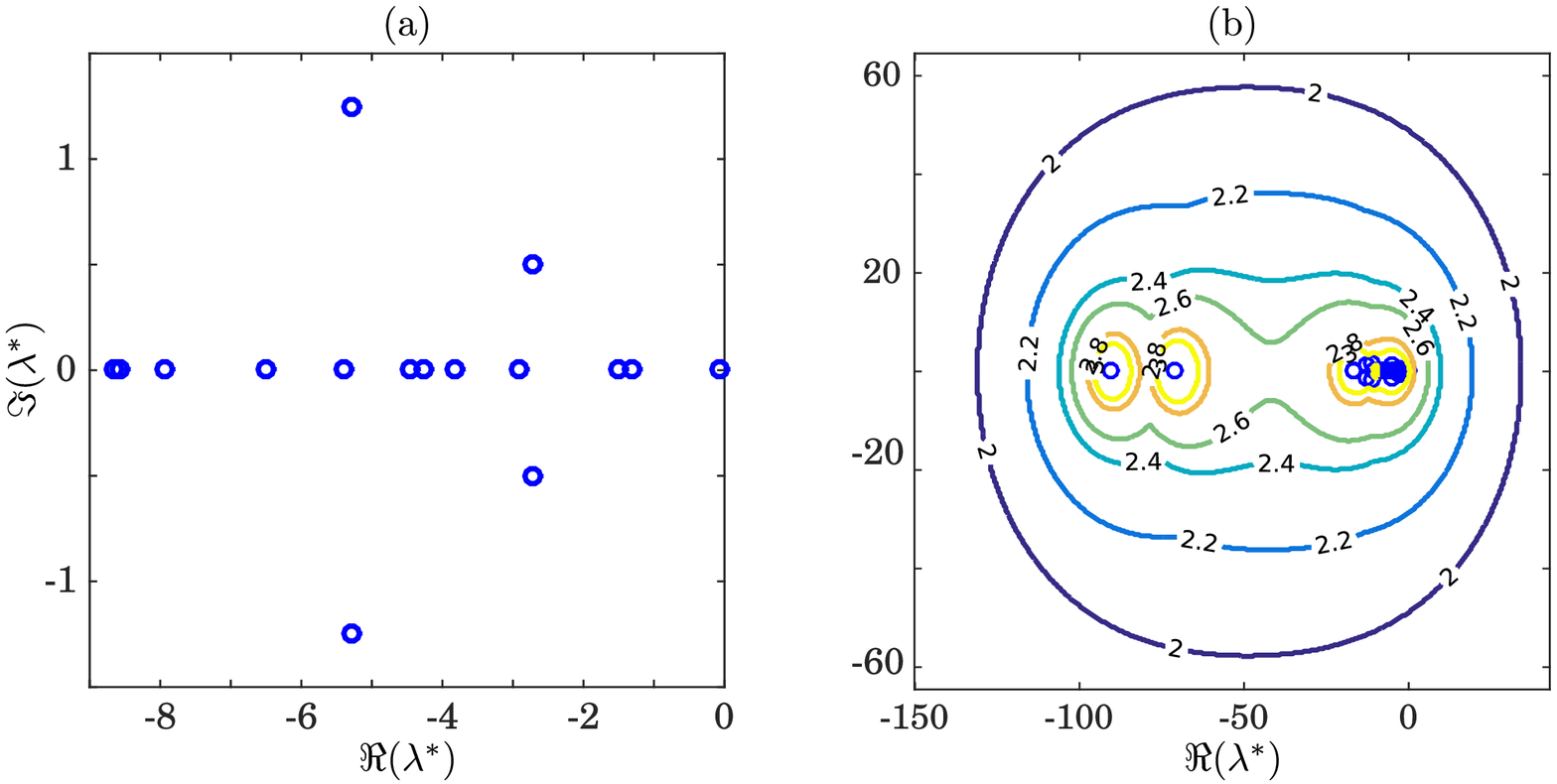}}
  \caption{(a) Eigenvalues of $\mrm{L}$ nondimenzionalized using the advective time scale $\tau_{adv}$. (b) The $\epsilon$-pseudospectrum of $\mrm{L}$ (Eqn.~\ref{eqn:Spectral2}) calculated following \citet{trefethen2005spectra}, which shows the non-normality of $\mrm{L}$. The numbers on the isolines are $-\log(\epsilon)$. In panel (a), only the first 15 eigenvalues (out of 25) are shown for better illustration and to keep the focus on the eigenvalues corresponding to the slowest-decaying modes. This has caused the large difference in the range of $x$-axis of the two panels.}
\label{fig:EVs}
\end{figure}

As shown in previous sections, the $\mrm{L}$ obtained using the GRF method can predict the time-mean response of the 3D RB system to external forcings, or the forcing needed for a given change in the time-mean flow, with high accuracy. In the present section, we study some of the spectral properties of $\mrm{L}$. Figures \ref{fig:Modes} and \ref{fig:EVs} show the four slowest-decaying eigenvectors of $\mrm{L}$ and its eigenvalues, respectively. 

The slowest-decaying mode is real, mostly in the interior (outside the boundary layers), and decays with a timescale of $\sim 17~\tau_{adv}$. This eigenvector coincides with $\mrm{L}$'s ``neutral vector'', which is the right singular vector with smallest singular number and the system's  most excitable dynamical mode because it is the largest time-mean response to external forcings \citep{Marshall1993,Goodman2002}. The leading POD of a turbulent flow (POD1) is expected to be identical to its neutral vector if the forcing from turbulent eddies is spatially uncorrelated and has uniform variance everywhere \citep{Goodman2002}. Figure~\ref{fig:Modes}a shows that the POD1 of the (unforced) DNS and $\mrm{L}$'s neutral vector are different, which is not surprising given that the presence of the boundary layers and turbulent plumes makes the flow anisotropic and spatially correlated. Just to demonstrate this point, for the $\mrm{L}$ calculated using the GRF method and Gaussian white noise $\boldsymbol{\xi}(t)$, we have integrated            
\begin{equation}
	 \dot{\boldsymbol{x}}(t) = \mrm{L} \boldsymbol{x}(t) + \boldsymbol{\xi}(t) \, , \label{eqn:Spectral1}
\end{equation}
using the Euler-Maruyama method. The leading POD of this dataset is shown in Fig.~\ref{fig:Modes}a, which, unlike the POD1 of DNS, agrees with the neutral vector of $\mrm{L}$.

The second slowest-decaying mode (Fig.~\ref{fig:Modes}b) is real as well but spans both the interior and boundary layers and decays faster than $1/\tau_{adv}$. The third slowest-decaying mode (Fig.~\ref{fig:Modes}c) is real and mostly varies in the interior. The fourth slowest-decaying mode (Fig.~\ref{fig:Modes}d) is complex with both real and imaginary parts of the eigenvector changing across the interior and boundary layers. This mode decays with the time scale of $\sim 0.37~\tau_{adv}$ and oscillates with the frequency of around $2 \tilde{\omega}$.    

Figure~\ref{fig:EVs}a shows the eigenvalues of $\mrm{L}$, which all have negative real parts (i.e., decaying). Except for the slowest-decaying mode, all other eigenmodes decay with timescales faster than $\tau_{adv}$; all eigenmodes of $\mrm{L}$ decay faster than the diffusive time scale $\tau_{diff}$ ($\sim 840 \tau_{adv} $). Figure~\ref{fig:EVs}b depicts the $\epsilon$-pseudospectrum of $\mrm{L}$ ($\Lambda_{\epsilon} (\mrm{L})$) given by \citep{trefethen2005spectra}
\begin{equation}
	 \Lambda_{\epsilon} (\mrm{L}) = \{ z \in C: \norm{(z\mrm{I} - \mrm{L})^{-1}}_2 \geq \epsilon^{-1} \} \, . \label{eqn:Spectral2}
\end{equation}
Here $\epsilon$ is the measure of proximity of a point in the complex plane $C$ to the spectrum of $\mrm{L}$. The calculated pseudospectrum shows that $\mrm{L}$ is non-normal and supports transient growth \citep{farrell1996generalized,trefethen1993hydrodynamic}. The non-normality of $\mrm{L}$ also suggests that estimating $\mrm{L}$ accurately using data-driven techniques such as Fluctuation-Dissipation Theorem (FDT) can face similar challenges reported in \citet{hassanzadeh2014wall} if POD modes are used as basis functions. In fact, recently \citet{Khodkar2018} have shown that for this system, POD-based FDT does not provide an accurate $\mrm{L}$. Guided by the results of Fig.~\ref{fig:EVs}(b), they proposed using DMD modes as the basis functions instead, and showed that DMD-enhanced FDT provides an accurate $\mrm{L}$, as accurate as the GRF-based LRF, for this system.

\color{black}

\section{\textrm{VIII}. Conclusions}\label{section:Conclusion}

We have developed a 1D linear ROM in the form of Eqn.~(\ref{eqn:Theory10}) for a 3D Rayleigh-B\'enard (RB) convection system, which is a fitting prototype for buoyancy-driven turbulence in various natural and engineering flows. Using the Green's function (GRF) method, we have calculated the LRF, $\mrm{L}$, and EFM, $\mrm{E}$, at $\mathrm{Ra}=10^6$. The EFM, $\mrm{E}$, is basically a matrix that parametrizes changes in the divergence of vertical eddy heat flux based on changes in the temperature profile. In section V, using several tests at $\mathrm{Ra}=10^6$, we have shown that $\mrm{L}$ and $\mrm{E}$ can accurately predict the time-mean responses of temperature and eddy heat flux to external forcings, and that  $\mrm{L}$ can well predict the forcing needed to change the mean flow in a specified way (inverse problem). Furthermore, we have shown in section VI that once these $\mrm{L}$ and $\mrm{E}$ are simply scaled by $\sqrt{\mathrm{Ra}/10^6}$, they work equally well for flows at other $\mathrm{Ra}$, at least in the investigated range of $ 5 \times 10^5 \le \mathrm{Ra} \le 1.25 \times 10^6$. 

The GRF method can be readily extended to use forcings that vary in the horizontal directions (e.g., applied at different Fourier modes one at a time) and time-dependent (e.g., applied at different frequencies one at a time). Such 3D ROMs, while computationally more expensive to calculate, can provide further insight into the spatio-temporal characteristics of buoyancy-driven turbulence.

The GRF method shows a promising performance for high-$\mathrm{Ra}$ turbulence, however, there are two issues that should be highlighted. First, a key assumption in developing the 1D ROM is linearity of the response. While it has been shown that at least for the large-scale atmospheric turbulence, $\mrm{L}$ and $\mrm{E}$ work well for responses/forcings that are large enough to be useful for various practical purposes \citep{hassanzadeh2015blocking,Hassanzadeh2016a,hassanzadeh2018quantifying,ma2017quantifying}, the limitations of the linearity assumption for the RB system and other problems should be explored in future studies. Second, the GRF method is computationally demanding because of the need for many forced full-dimensional simulations (although, these simulations are needed only once, e.g., for the purpose of online flow control/optimization, the calculations can be done offline and the calculated LRF can then be used online with negligible computational cost). While the simple scaling found here suggests that the LRF and EFM do not have to be calculated for every Rayleigh number (at least for a range of $\mathrm{Ra}$), the numerical cost can limit its use as a generally applicable method (particularly to build 3D ROMs). Still, calculating the accurate 1D and 3D ROMs using the GRF method for some turbulent systems has the following major advantages:
\begin{enumerate}
\item Knowing the accurate $\mrm{L}$ can guide developing better data-driven techniques, as for example, done in \citet{Hassanzadeh2016b}. In particular, comparing the flow's Koopman/DMD modes with the eigen/singular vectors of the $\mrm{L}$ calculated here might be informative. In another direction, while we have not attempted to optimize the basis functions used in the GRF method in this work, the Koopman/DMD modes might provide some insight into better/optimal basis functions for the GRF method, which can reduce the computational cost and improve the accuracy. 
\item Analyzing the spectral properties of $\mrm{E}$ can help with better understanding the physics of eddy fluxes and improving the turbulence closure schemes, which connects with the ongoing efforts in developing better deterministic and stochastic parameterizations for geophysical turbulence \citep{Cuba2011, cooper2015optimisation, Benosman2017, tan2018extended}.    
\end{enumerate}
The authors aim to follow these lines of research in their future work. 
 
\color{black}

\section*{Acknowledgment \label{section:Acknowledgment}}

We thank Thanos Antoulas and Matthias Heinkenschloss for fruitful discussions, Arthi Appathurai for help with conducting some of the simulations, and two anonymous reviewers for insightful comments. This work was supported by funding  provided to P.H. by the NASA grant 80NSSC17K0266, the NSF grant AGS-1552385, a Faculty Initiative Fund award from the Rice University Creative Ventures, and the Mitsubishi Electric Research Labs. This work used the Extreme Science and Engineering Discovery Environment (XSEDE) Stampede2 through allocation ATM170020, the Yellowstone high-performance computing system provided by NCAR's Computational and Information Systems Laboratory through allocation NCAR0462, and the DAVinCI cluster of the Rice University Center for Research Computing.

\color{black}




\bibliography{Main}

\end{document}